\documentclass{article}
\usepackage{graphicx}
\usepackage{titlesec}
\usepackage{braket}
\title{Proceedings of the talk "Study of CP violation in charm meson decays"\\ for XXX Epiphany conference (Krakow, Jan 8-11 2024)}
\author{Aleksei Chernov (aleksei.chernov@cern.ch), \\ Institute of Nuclear Physics PAN, Krakow, Poland}

\date{}

\titleformat{\section}
{\normalfont\fontsize{12}{14}\bfseries}{\thesection}{1em}{}
\titleformat{\subsection}
{\normalfont\fontsize{12}{14}\bfseries}{\thesubsection}{2em}{}

\begin{document}
\maketitle

\begin{abstract}
	\noindent
	\small
	Basic idea of $CP$ violation and its importance for cosmology, Standard Model and possible New Physics are introduced. Recent (2019) discovery of CP violation in charm in the $\Delta A_{CP}$ between $D^0\rightarrow K^-K^+$ and $D^0\rightarrow \pi^+\pi^-$ decay channels and evidence for direct CP violation in $D^0\rightarrow \pi^-\pi^+$ decays in 2023 is discussed. Motivation and brief outline for search of CP violation in the $D^0\rightarrow V\gamma$, where $V=\phi,\rho^0,
	\bar{K}^{*0}$ decays is given.
\end{abstract}
\section{Introduction}
$CP$ (charge-parity conjugation) is the operator that changes a particle $X$ into its anti-particle $\bar{X}$ and reverses spatial coordinates. Non-conservation of $CP$ symmetry (henceforth $CP$ violation) is the second of Sakharov's cosmological conditions \cite{Sakharov - cosmology}, necessary to explain observed prevalence of baryons over anti-baryons in the observable Universe. In the Standard Model (SM), the only measurable source of $CP$ violation is a weak phase in the quark mixing CKM matrix. Because CP violation in SM is small, some New Physics (NP) is believed to be necessary in order to explain predominance of baryons (\cite{Charm CP theory}).\newline
One can distinguish between three types of $CP$ violation: $CP$ violation in the decay, $CP$ violation in the mixing of neutral mesons (e.g. $D^0$ and $\bar{D}^0$), and $CP$ violation in the interference between mixing and decay.
CP violation in decay occurs when partial decay widths for final states $f$ and $\bar{f}$ are different for $D^0$ and $\bar{D}^0$ flavours:
\begin{equation}
	A_{CP} = \frac{\Gamma(D^0\rightarrow f)-\Gamma(\bar{D}^0\rightarrow\bar{f})}{\Gamma(D^0\rightarrow f)+\Gamma(\bar{D}^0\rightarrow\bar{f})}
	\label{Asymmetry_definition_theory}
\end{equation}
In experimental terms, measuring CP asymmetry usually means measuing raw asymmetry by counting flavour $tags$ (pions for $D^{*+}\rightarrow D^0\pi^+$ and muons for $B^+\rightarrow \bar{D}^0\mu^+{\nu_\mu}$, h.c. implied) and substracting $nuisance$ asymmetries, e.g production and detection asymmetries:
\begin{equation}
	A_{raw} = \frac{N_{D^0}-N_{\bar{D}^0}}{N_{D^0}+N_{\bar{D}^0}}\approx A_{CP}+A_{det.}+A_{prod.}
	\label{Asymmetry_definition_experiment}
\end{equation}
More about those additional sources of observed asymmetry will be said in Section \ref{sec-nuisance}.
These proceedings are focused on the CP violation in decays. CPV in $D^0-\bar{D}^0$ mixing results in a time-dependent correction to the measured direct asymmetry:
	$A_{CP}(f) \approx a^{d}_f + \frac{\braket{t}_f}{\tau_D}\Delta Y_f$;
where $t_f$ is mean decay time for $D^0$ in the experimental data, $\tau_D$ is mean lifetime of $D^0$ and $\Delta Y_f$ is one of the parameters quantifying CPV in mixing and in interference between mixing and decay, previously measured at B-factories (\cite{BaBaR_D0_mixing},\cite{Belle_D0_mixing}) and LHCb (\cite{LHCb_D0_mixing}).
\subsection{CP violation in the decays of charmed mesons}
\begin{figure}[hbt]
	\centering
	\includegraphics[width=0.4\textwidth]{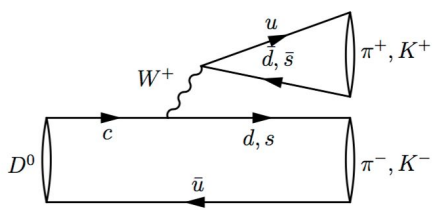}
	\includegraphics[width=0.4\textwidth]{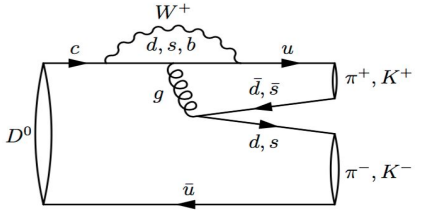}
	\caption{\small Example of a tree-level and one-loop (penguin) diagrams of \newline$D^0\rightarrow K^-K^+$ and $D^0\rightarrow\pi^-\pi^+$ decays. Weak CKM phase enters through the penguin $c\rightarrow u$ transition.}
	\label{d2hh_cartoon}
\end{figure}

Within the Standard Model, $CP$ violation in the charm decays is expected to be small - $10^{-4}\div10^{-3}$ (see, for example, \cite{Charm CP theory}), owing to suppressed penguin amplitude compared to the tree level due to small differences between masses of light quarks ($s,d$) and strong CKM suppression of the $V_{ub}$ coupling. However, contribution of some NP to the penguin amplitude (Fig.\ref{d2hh_cartoon}, right) can alter the $A_{CP}$ compared to SM predictions. 
Although it was predicted for a long time, the first observation of the CP violation in charm was reported by the LHCb experiment in 2019 \cite{DeltaA_CP_Run2}. That analysis utilized $\pi^\pm$-tagged charm decays: $D^{*+}\rightarrow(D^0\rightarrow h^- h^+)\pi^+_{tag}$, where $h^+ = K^+,\pi^+$ (here and afterwards charge conjugation is implied unless stated otherwise), promply produced in $pp$ collisions, as well as muon-tagged $D^0\rightarrow h^-h^+$ decays produced in semileptonic $\bar{B}\rightarrow D^0\mu^-\bar{\nu}_\mu X$ decays, for $5.9\:fb^{-1}$ of Run 2 data. Combined with the previous LHCb results \cite{DeltaA_CP_Run1}, it gives first observation of $CP$ violation in decays of charmed mesons at more than 5$\sigma$:
$\Delta A_{CP} \equiv A_{CP}(D^0\rightarrow K^-K^+) - A_{CP}(D^0\rightarrow\pi^-\pi^+) = (-15.4\pm2.9)\times10^{-4}$
, where error contains both statistical and systematic contributions (but is dominated by statistics).\newline
In order to obtain such extraordinary precision in $A_{CP}$ asymmetries for individiual decay channels, it is necessary to control for other sources of asymmetry, which will be discussed in the next section. The nuisance asymmetries cancel out within the $\Delta A_{CP}$, but the price to pay for measuring $\Delta A_{CP}$ is the inability to disentangle invidiual asymmetries. 
\section{Direct CP violation in $D^0\rightarrow h^+h^-$ decays}
\subsection{Nuisance asymmetries}
\label{sec-nuisance}
Measured ($raw$) asymmetry can be driven not only by CP violation. The main sources of nuisance asymmetries at the LHCb are production ($\sigma(pp\rightarrow D^{*+})\neq\sigma(pp\rightarrow{D}^{*-})$) and detection asymmetry  (Eq.\ref{Asymmetry_definition_experiment}) between positively and negatively charged hadrons - latter arises due to different detection efficiency. In the busy environment of a hadron collider, it is considered implausible to precisely calculate production and detection asymmetries for any given decay from the theory and/or from simulation, and therefore, a data-driven approach is required. A reference channel is selected, with the final state similar to a studied decay, and asymmetry is measured for that channel.\newline
An important caveat is that $A_{det}$ and $A_{prod}$ are dependent on kinematics, and the asymmetry tends to be larger for smaller momenta. 
Therefore, it is necessary to use kinematics matching procedure - usually done by assigning weights on per-event based until distributions of kinematic variables $(p,\eta,\phi\:(\textit{azimuthal angle}))$ match between corresponding particles in signal and reference data. This usually leads to a significant drop in statistics of weighted dataset compared to unweighted one. Multiple reference channels can be used in order to mitigate that penalty. 
\subsection{Measurement of $A_{CP}(D^0\rightarrow K^-K^+)$}
\begin{figure}[hbt]
	\centering
	\includegraphics[width=0.475\textwidth]{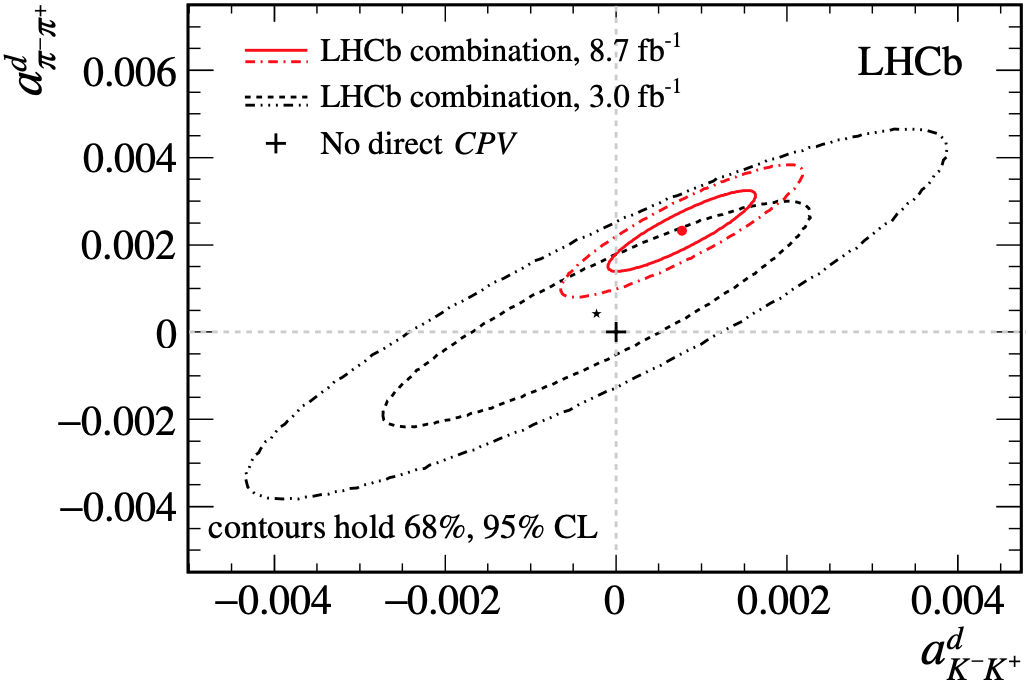}
	\includegraphics[width=0.475\textwidth]{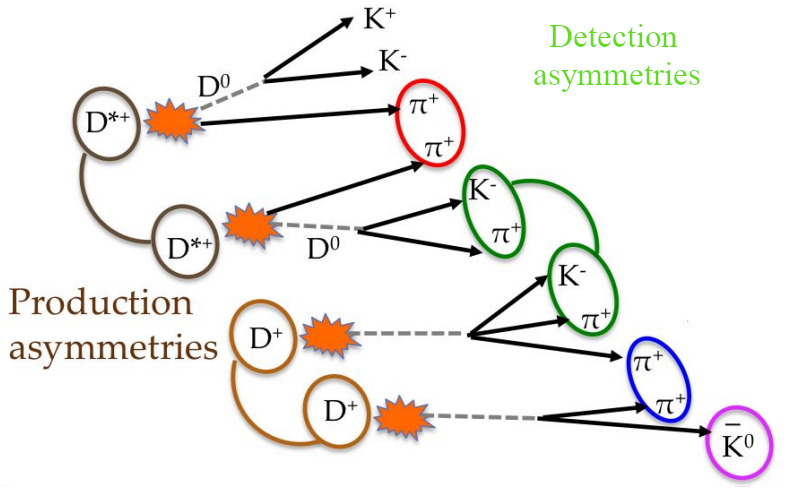}
	\caption{\small Left - Recent measurement of $A_{CP}(D^0\rightarrow h^-h^+)$ by LHCb shows tension with the CP conservation. Right - illustration of calibration with $D^+$ decays used to extract $A_{CP}$ from measured $A_{raw}$}.
	\label{Beatiful D2hh results}
\end{figure}

After discovering the CP violation in the charm sector, the next step is to disentangle individual asymmetries $A_{CP}(D^0\rightarrow K^-K^+)$ and $A_{CP}(D^0\rightarrow \pi^-\pi^+)$ from $\Delta A_{CP}$. Two complementary calibration procedures are introduced: through  $D^+\rightarrow K^-\pi^+\pi^+$, $D^+\rightarrow\bar{K}^0\pi^+$, and $D^+_s\rightarrow\phi\pi^+$,$\:D^+_s\rightarrow\bar{K}^0K^+$ decays, and, with either, $D^0\rightarrow K^-\pi^+$. All those are high-statistics and high-purity channels dominated by Cabbibo-favoured transitions with the expectation of no CP violation in the SM.
The explicit formula for each calibration follows, and Fig.\ref{Beatiful D2hh results} (right) demonstrates cancellation of nuisance asymmetries:
{
	\nopagebreak
	\small
\begin{eqnarray*}
A_{CP}(K^-K^+)=A_{raw}(D^0\rightarrow K^-K^+)-{A_{raw}(D^0\rightarrow K^-\pi^+)}+\\+A_{raw}(D^+\rightarrow K^-\pi^+\pi^+)-A_{raw}(D^+\rightarrow\bar{K}^0\pi^+) + A(\bar{K}^0)
\end{eqnarray*}
\vspace{-0.75cm}
\begin{eqnarray*}
	A_{CP}(K^-K^+) = A_{raw}(D^0\rightarrow K^-K^+)-A_{raw}(D^0\rightarrow K^-\pi^+)+\\+A_{raw}(D^+_s\rightarrow \phi\pi^+)-A_{raw}(D^+_s\rightarrow\bar{K}^0K^+)+A(\bar{K}^0)
\end{eqnarray*}
}

 \noindent Last term arises from combined effects of $CP$ violation and mixing in the neutral kaon system and different interaction rates of $K^0$ and $\bar{K}^0$ with the detector material. After combining the results from both, the most precise measurement of $A_{CP}(D^0\rightarrow K^-K^+)$ was obtained \cite{D2KK_DeltaA_CP_Run2}: $A_{CP}(D^0\rightarrow K^-K^+) = (6.8\pm5.4(stat.)\pm1.6(sys.) ) \times 10^{-4}$.
By itself, it is quite consistent with hypothesis of no CP violation. However, after substracting time-dependent asymmetry due to $D^0-\bar{D}^0$ mixing and combining this result with the $\Delta A_{CP}$ analysis \cite{DeltaA_CP_Run2}, one obtains:
$a^{d}_{CP}(\pi^-\pi^+) = 23.2\pm6.1\times10^{-4}\:(3.8\sigma)$, $a^{d}_{CP}(K^-K^+)=7.7\pm5.4\times10^{-4}$. This is the first evidence for a direct CPV in an individual decay. Fig. \ref{Beatiful D2hh results} (left) shows a comparison between results based on Run 1 and Run 1+Run 2 LHCb data.\\
In the limit of the $SU(3)_F$ symmetry for $m_u=m_d=m_s$, $a^{d}_{CP}(K^-K^+)=-a^{d}_{CP}(\pi^-\pi^+)$ \cite{Grossman CPV theory}. Breaking $SU(3)_F$ leads to the different values, with exact size of the effect being uncertain. The experimental results are  in disagreement with those predictions. 
\section{Radiative charm decays $D^0\rightarrow V\gamma$}
Decays $D^0\rightarrow V\gamma$, where $V=\phi,\rho^0,\bar{K}^{*0}$, were studied at B-factories \cite{Belle PhiGamma}, with observed branching ratios of $10^{-5}\div10^{-4}$ and $A_{CP}$ consistent with null at $\%$ level uncertainties.
\begin{figure}[b]
	\includegraphics[width=0.32\textwidth]{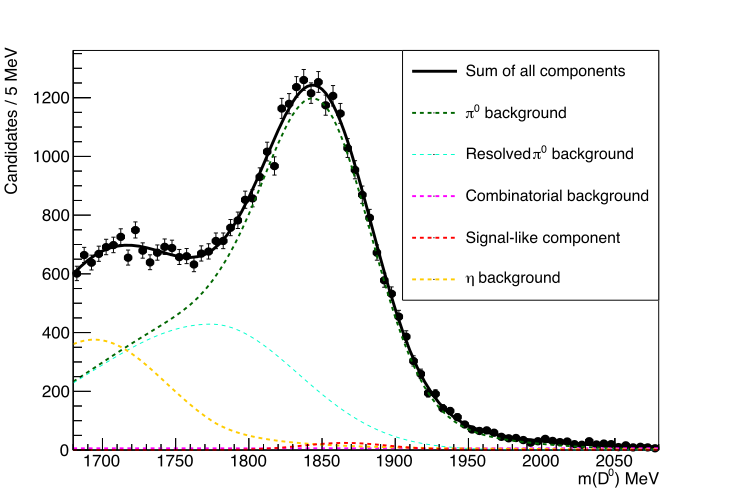}
	\includegraphics[width=0.32\textwidth]{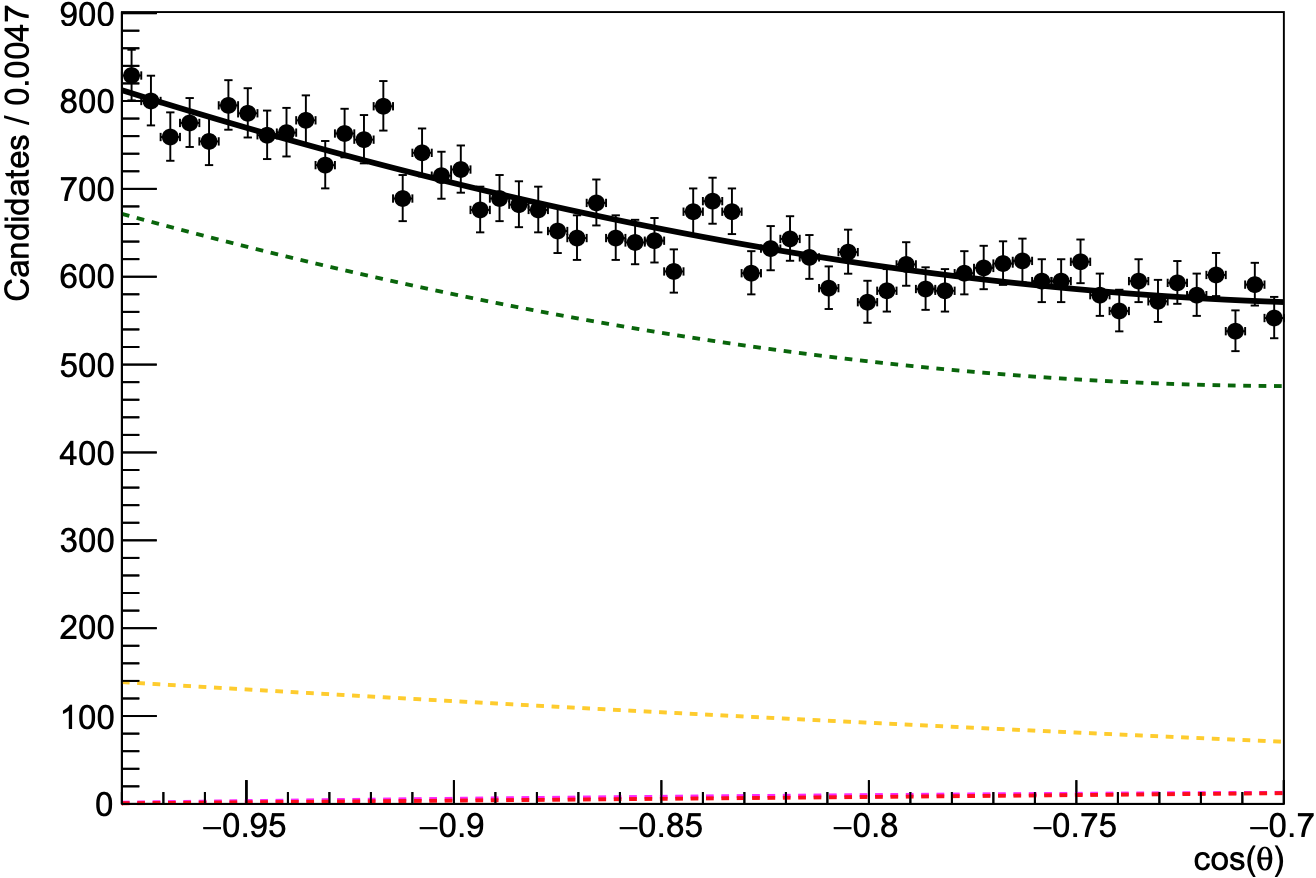}
	\includegraphics[width=0.32\textwidth]{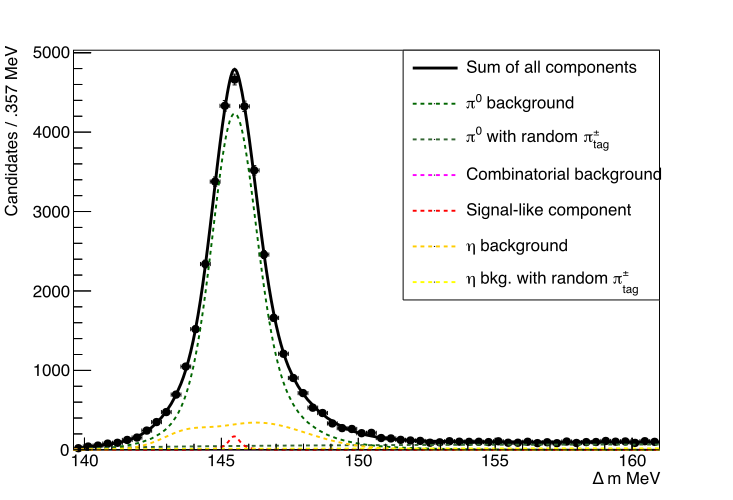}
	\caption{\small Distributions of $m(D^0), \cos{\theta}$ and $\Delta m$ for $D^0 \rightarrow \bar{K}^{*0} \gamma$ pseudodata in signal-suppressed $\cos\theta$ region. Results of three-dimensional fit are superimposed. Fit components are specified in the legend. Number of events generated is 40000.}
	\label{KstGamma toys hel edge}
\end{figure}
SM predicts CP violation of up to $\sim10^{-3}$ for $\phi$ and $\rho^0$, which can be enhanced further by NP \cite{Radiative charm theory}. Studying $D^0\rightarrow V\gamma$ is possible at LHCb, but there are challenges to overcome. Irreducible peaking background from $D^0\rightarrow V\pi^0$ - decays with branching ratio $\sim 10^{-3}$ - one or two orders of magnitude over the signal, and boosted $\pi^0\rightarrow\gamma\gamma$ are usually reconstructed as a single energy cluster in the LHCb electromagnetic calorimeter. \\
We aim to measure $A_{CP}(D^0\rightarrow V\gamma)$.  To that end, LHCb analysis employs $D^0 \rightarrow K^-K^+$ and $D^0\rightarrow \pi^-\pi^+$ reference channels to correct for nuisance asymmetries in $D^0\rightarrow(\phi\rightarrow K^-K^+)\gamma$ and $D^0\rightarrow(\rho^0\rightarrow \pi^-\pi^+) \gamma$, respectively, whereas $D^0 \rightarrow K^-\pi^+\pi^0$ is used for $D^0\rightarrow (\bar{K}^{*0}\rightarrow K^-\pi^+ )\gamma$. We can measure: $\Delta A_{CP}^{rad.}=A_{CP}(D^0 \rightarrow V \gamma) -  A_{CP}( D^0 \rightarrow h^-h^+(\pi^0) ) \approx A_{raw}(D^0 \to V \gamma) - A_{raw}( D^0 \to h^-h^+ (\pi^0) )$, after the kinematics matching, as discussed in Section 2.1, then take measured $D^0\rightarrow h^-h^+$ asymmetries (\cite{D2KK_DeltaA_CP_Run2},\cite{DeltaA_CP_Run2}) as external inputs in order to obtain individiual $A_{CP}(D^0\rightarrow V\gamma)$.\\
A multivariate tool IsPhoton, based on shapes of energy clusters in the calorimeter, is used for $\gamma/ \pi^0$ separation. Signal is distinguished from residual background using three variables: $D^0$ invariant mass $m(D^0)$, the difference between $D^{*+}$ and $D^0$ masses $\Delta m = m(D^{*+})- m(D^0)$, and $\cos{\theta}$. Helicity angle of the $V$ meson $cos\theta$ is important, as its' expected distribution for $D^0\rightarrow V\pi^0$ is similar to $\cos^2 {\theta}$, whilst for signal channels $D^0\rightarrow V\gamma$ it is $\sin^2\theta$. A three-dimensional maximum likelihood fit is performed to these variables, taking into account correlations between $m(D^0)$ and $\Delta m$. Figures \ref{KstGamma toys hel edge} and \ref{KstGamma toys} show $m(D^0)$, $\cos\theta$ and $\Delta m$ distributions for samples of pseudodata based on real $D^0 \rightarrow \bar{K}^{*0}\gamma$ decays reconstructed in the Run 1 LHCb data. Number of psuedo-events generated is similar to number of events observed in real data after selection. These are combined $\pi_{tag}^+$ and $\pi_{tag}^-$ samples, and fit results are superimposed. Fig. 3 shows the sample for the $\cos\theta$ region with negligible signal contribution. It is used for calibration of the $\pi^0$ background description. Sample for the signal-enhanced $\cos\theta$ region is presented in Fig. 4. 
	\begin{figure}[hb]
	\includegraphics[width=0.3\textwidth]{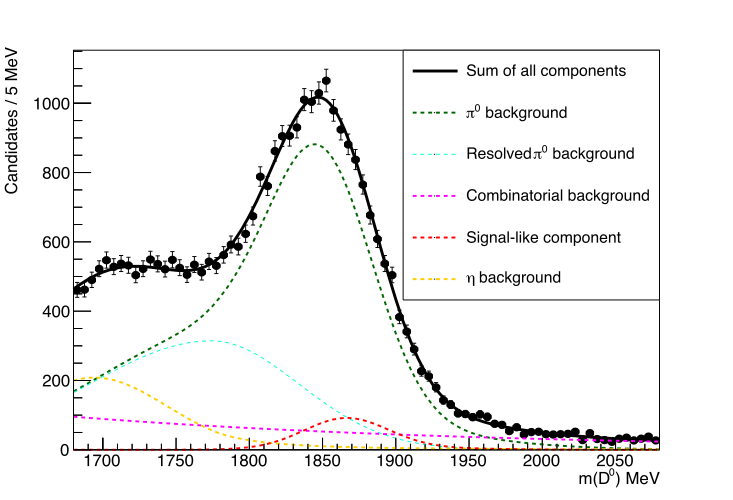}
	\includegraphics[width=0.3\textwidth]{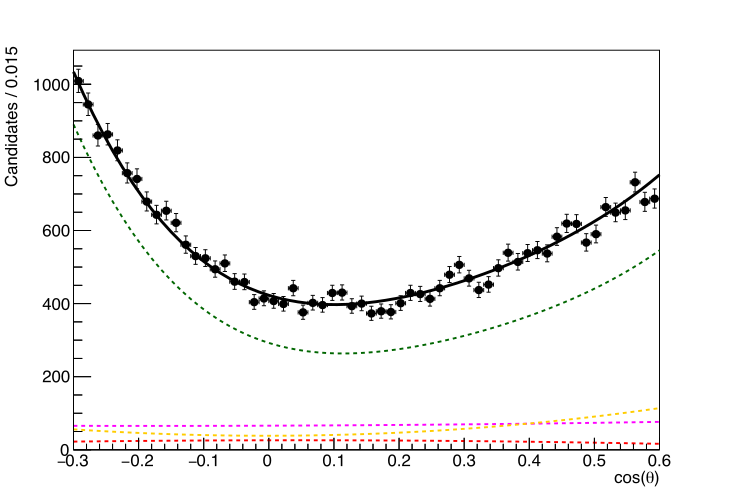}
	\includegraphics[width=0.3\textwidth]{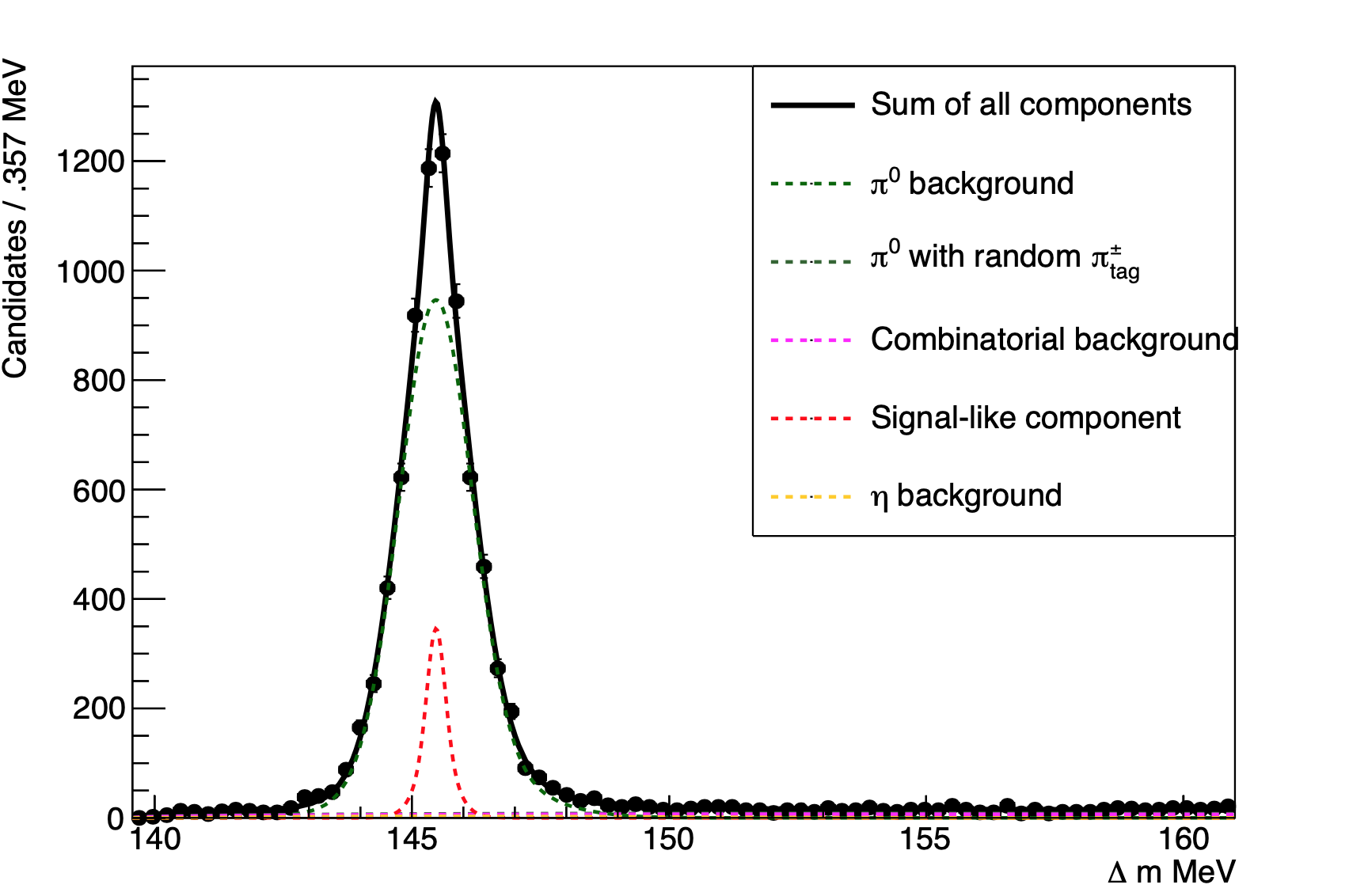}
	\caption{
		\small Distributions of $m(D^0), \cos\theta$ and $\Delta m$ for $D^0 \rightarrow \bar{K}^{*0} \gamma$ pseudodata in signal-enhanced $\cos\theta$ region. Results of three-dimensional fit are superimposed. Fit components are specified in the legend. Number of events generated is 33000.}	
	\vspace{-0.4cm} 
	\label{KstGamma toys}
\end{figure}
\section{Conclusion and outlook}
Direct $CP$ violation in charm sector had been observed through $\Delta A_{CP}$ in 2019 \cite{DeltaA_CP_Run2}. A follow-up analysis \cite{D2KK_DeltaA_CP_Run2} shows evidence of direct CP violation in the individual decay channel $D^0\rightarrow \pi^-\pi^+$. When combined, these two results demonstrate some tension with theoretical prediction, although difficulty of precise theoretical calculations in charm \cite{Charm CP theory, Grossman CPV theory} makes it hard to judge whether this is a sign of NP. Other decay modes, such as radiative charm decays $D^0\rightarrow V\gamma$ can provide more insight into $CP$ violation in charm.
{\\ \\ \noindent \bf \small Acknowledgements.} I would like to express my gratitude to the National Science Centre NCN in Poland, for financial support under the contract no. 2017/26/E/ST2/00934.


\begin{thebibliography}{40}
\small
\vspace{-0.25cm}
\bibitem{Sakharov - cosmology} A.Sakharov, 
JETP Lett. 5,24-27 (1967).
\vspace{-0.1cm}
\bibitem{Charm CP theory} Avital Dery, Yosef Nir, 
JHEP12, 104 (2019) .
\vspace{-0.1cm}
\bibitem{BaBaR_D0_mixing} J. P. Lees et al. (BABAR Collab.), 
Phys. Rev. D 87, 012004 (2013).
\vspace{-0.1cm}
\bibitem{Belle_D0_mixing} T. Peng et al. (Belle Collab.),
Phys. Rev. D 89, 091103 (2014).
\vspace{-0.1cm}
\bibitem{LHCb_D0_mixing} R. Aaij et al. (LHCb Collab.), 
Phys. Rev. Lett. 127, 111801 (2021)
\vspace{-0.1cm}
\bibitem{DeltaA_CP_Run2} R.~Aaij et al. (LHCb Collab.), 
Phys. Rev. Lett. 122, 211803 (2019). 
\vspace{-0.1cm}
\bibitem{D2KK_DeltaA_CP_Run2} R.~Aaij et al. (LHCb Collab.), 
Phys. Rev. Lett. 131, 091802 (2023).
\vspace{-0.1cm}
\bibitem{DeltaA_CP_Run1} R. ~Aaij et al. (LHCb Collab.), 
Phys. Rev. Lett. 116 (2016) 191601.
\vspace{-0.1cm}
\bibitem{Belle PhiGamma} T. Nanut et al.(Belle Collab.),
PRL 118, 051801 (2017).
\vspace{-0.1cm}
\bibitem{Radiative charm theory} S. de Boer and G. Hiller, 
JHEP 08 (2017) 091.
\vspace{-0.1cm}
\bibitem{Grossman CPV theory} Y. Grossman, A. L. Kagan, and Y. Nir, 
Phys. Rev. D75 (2007) 036008.
\end{thebibliography}
\end{document}